\documentclass[a4paper]{article}
\usepackage{CJK}
\usepackage[paperwidth=185mm,paperheight=230mm,textheight=190mm,textwidth=145mm,left=20mm,right=20mm, top=25mm, bottom=20mm]{geometry}
\usepackage[CJKbookmarks, colorlinks, bookmarksnumbered=true,pdfstartview=FitH,linkcolor=blue,citecolor=green]{hyperref}
\usepackage{amsmath,amssymb}
\usepackage{amsthm}
\usepackage{calc}
\usepackage{graphicx}
\usepackage{supertabular}
\usepackage{longtable}
\usepackage{float}
\usepackage{color}
\usepackage{enumerate}
\usepackage{colortbl,booktabs}
\usepackage{natbib}
\usepackage{bm}
\usepackage{multirow}
\usepackage{authblk}

\def\wh{\widehat}
\def\wt{\widetilde}

\begin{document}

\title{ Kernel partial least squares regression for functional nonlinear models }
\author[a]{{\fontsize{12pt}{18pt}\selectfont Rou Zhong}}
\author[a]{{\fontsize{12pt}{0.5em}\selectfont Dongxue Wang}}
\author[a]{{\fontsize{12pt}{0.5em}\selectfont Jingxiao Zhang} \thanks{zhjxiaoruc@163.com}}
\affil[a]{{\emph\fontsize{12pt}{0.5em}\selectfont Center for Applied Statistics, School of Statistics, Renmin University of China}}
\date{}
\maketitle

\begin{abstract}

Functional regression is very crucial in functional data analysis and a linear relationship between scalar response and functional predictor is often assumed. However, the linear assumption may not hold in practice, which makes the methods for linear models invalid. To gain more flexibility, we focus on functional nonlinear models and aim to develop new method that requires no strict constraint on the nonlinear structure. Inspired by the idea of the kernel method in machine learning, we propose a kernel functional partial least squares (KFPLS) method for the functional nonlinear models. The innovative algorithm works on the prediction of the scalar response and is accompanied by R package \texttt{KFPLS} for implementation. The simulation study demonstrates the effectiveness of the proposed method for various types of nonlinear models. Moreover, the real world application also shows the superiority of the proposed KFPLS method.

\textbf{ Keywords }: Nonlinear models, functional regression, kernel technique, partial least squares
\end{abstract}

\section{ Introduction }

Functional regression plays a crucial role in functional data analysis (FDA), and it is quite useful when the considered model involves functional predictor or functional response. In this paper, we focus on models with functional predictors and scalar response, which is very common in practice and has received much attention in the existing work.

Among researches on functional regression, functional linear regression (FLR) model has been widely studied \citep{hall2007methodology, james2009functional, yuan2010reproducing, zhou2022functional}. Although the linearity assumption simplifies the estimation to a large extent, it cannot always be satisfied in real data application. Therefore, more and more concerns have been devoted to the study of functional nonlinear models. \citet{muller2008functional} relaxed the linearity assumption by extending the functional linear model to a functional additive model (FAM) with the functional principal component scores as predictors. Different from \citet{muller2008functional}, \citet{muller2013continuously} considered the additivity of the time domain rather than the spectral domain, and a continuously additive model was introduced. \citet{mclean2014functional} proposed a functional generalized additive model (FGAM), which can be seen as a functional extension of generalized additive models. Based on FAM in \citet{muller2008functional}, \citet{zhu2014structured} took into account the sparse structure of the additive components. \citet{fan2015functional} discussed the cases with multiple functional predictors. They presumed additivity of these functional predictors in the model and presented a functional additive regression method for such problem. Moreover, \citet{chen2011single} studied single and multiple index functional regression models with unknown link functions. \citet{huang2022estimation} further developed method for functional single index models to achieve optimal convergence rate of the slope function estimator.

According to the existing works, difficulties in handling functional nonlinear models can be summarized in two aspects. First, functional data are infinite dimensional, which makes the use of fully nonparametric methods unsuitable and makes curse of dimensionality have to be considered. Most of the developed methods focused on models with one functional predictor, and it can be more complex when multiple functional predictors occur. Second, various types of structures have been imposed on the nonlinear relationship between predictors and response, such as additive structure and single index structure, to make the estimation feasible. As the assumed structures may not be consistent with real world data, developing approaches for models with less restrictive structures can be quite beneficial.

In this paper, we propose a kernel functional partial least squares (KFPLS) method, which can overcome the above mentioned difficulties, for functional nonlinear models. We borrow the idea of the kernel method \citep{hofmann2008kernel}, which is a common technique in machine learning, and \citet{song2021nonlinear} also considered the kernel method in functional principal component analysis (FPCA). Specifically, for our work, we assume that the response and predictors can be linearly related after an unknown nonlinear map of the multiple functional predictors. We further demonstrate that this is a fairly general assumption. Then we can obtain a linear model which can be solved by the partial least squares (PLS) method. In specific, the nonlinear iterative partial least squares (NIPALS) algorithm is applied \citep{rosipal2001kernel, rosipal2011nonlinear}. Although the mapped functional predictors are unknown, only inner products involve in the computation. Therefore, kernel trick can be employed and the inner products are replaced by the corresponding values of kernel function. Moreover, the number of components of PLS and the tuning parameters of kernel function are selected by cross-validation (CV). It is shown in our simulation study that KFPLS method performs effectively regardless of whether the data are generated through linear model or nonlinear models.

The contributions of this paper are three-fold. First, the proposed KFPLS method is more flexible, since only mild conditions are needed. In the simulation study, we consider different types of functional nonlinear models. And our approach shows advantage when compared with other methods, which indicates KFPLS method can adapt to more kinds of structures of data. Second, to the best of our knowledge, the idea of kernel method has not been utilized in functional regression analysis before. Thus, our work provide a complete new estimating procedure for functional nonlinear models. Third, the developed algorithm is comprehensible and can be implemented easily. Moreover, the \textsf{R} package \texttt{KFPLS} can be downloaded from \url{https://CRAN.R-project.org/package=KFPLS}.

The rest of the paper is organized as follows. In Section \ref{SecMethod}, we introduce our assumption of the functional nonlinear model and present the proposed KFPLS methods. The numerical performance of our method is assessed by the comparison with other approaches in Section \ref{SecSim}. The results of the real data analysis for the Tecator dataset are reported in Section \ref{SecReal}. In Section \ref{SecDis}, we finally conclude this paper with some discussions.

\section{ Methodology }\label{SecMethod}

\subsection{Functional nonlinear model}

Suppose that there are $n$ independently and identically distributed subjects in the study. For the $i$-th subject, a scalar response $Y_i$ and $p$ functional predictors $X_{i1}, \ldots, X_{ip}$ are measured, where $X_{ij} \in \mathcal{X}$ for $j = 1, \ldots, p$, and $\mathcal{X}$ is a Hilbert space of functions that map from a bounded and closed interval $\mathcal{T}$ to $\mathbb{R}$. The functional nonlinear model is of the following form
\begin{align}
Y_i = f(X_{i1}, \ldots, X_{ip}) + \epsilon_i = f(\wt{X}_i) + \epsilon_i, i = 1, \ldots, n, \label{model_raw}
\end{align}
where $\wt{X}_i = (X_{i1}, \ldots, X_{ip})^\top \in \mathcal{X}^p$, $\epsilon_i$ is independent of $\wt{X}_i$, it is the measurement error with zero mean and finite variance, and $f$ is an unknown function that maps elements in $\mathcal{X}^p$ to $\mathbb{R}$. For a new observation of functional predictor, which is denoted as $\wt{X}_0$, we are interested in obtaining the estimation of $f(\wt{X}_0)$, which can also be seen as the prediction of $Y_0$.

However, it can be intractable to get $\wh{f}$ without any assumptions on $f$, since curse of dimensionality must be taken into account. That is the reason why additive structure and single index structure are often imposed on $f$. We aim to weaken these limits and develop new method that can accommodate to more general cases. Inspired by the idea of kernel method which is popular for nonlinear problems in machine learning, we assume that there exists a nonlinear map $\bm{\Phi}: \mathcal{X}^p \rightarrow \mathbb{R}^M$, such that $Y_i$ and $\bm{\Phi}(\wt{X}_i)$ can be linearly related, where $M$ is the dimension of the mapped data. Then the model can be rewritten as
\begin{align}
Y_i = \bm{\Phi}(\wt{X}_i)^\top \bm{\beta} + \epsilon_i, i = 1, \ldots, n, \label{model_lin}
\end{align}
where $\bm{\beta}$ is the coefficient.

More explanations of (\ref{model_lin}) are given here. First, $\wt{X}_i$ is mapped to $\mathbb{R}^M$ rather than a general dot product space which is called feature space in machine learning. By doing this, we can treat $\bm{\Phi}(\wt{X}_i)$ in a vector form, and approaches of linear model in multivariate analysis can directly applied to (\ref{model_lin}). The mapped space can be further extended to some general dot product space as long as the estimating method for (\ref{model_lin}) can be justified in the corresponding space. Second, the dimension $M$ can be finite or diverge to infinity, which depends on the kernel function that being used in the algorithm. Therefore, the map $\bm{\Phi}$ can in fact be seen as a nonlinear dimension reduction of the functional predictors, which is quite different from the intuition of kernel method in machine learning. This difference is not surprising since functional data are infinite dimensional, and it is reasonable to consider linear model after extracting effective information via nonlinear dimension reduction when handling (\ref{model_raw}). Third, (\ref{model_lin}) is very flexible as no strict constraint is required for the nonlinear map $\bm{\Phi}$. Hence, the additive models and single index models considered in the existing papers, such as \citet{muller2008functional} and \citet{chen2011single}, can be regarded as a special case of (\ref{model_lin}).

\subsection{Partial least squares}

We first expressed (\ref{model_lin}) in a matrix form, that is
\begin{align}
\textbf{Y} = \bm{\Psi} \bm{\beta} + \bm{\epsilon}, \label{model_mat}
\end{align}
where $\textbf{Y} = (Y_1, \ldots, Y_n)^\top$, $\bm{\epsilon} = (\epsilon_1, \ldots, \epsilon_n)^\top$, $\bm{\Psi} = (\Psi_1, \ldots, \Psi_n)^\top$ and $\Psi_i = \bm{\Phi}(\wt{X}_i)$. For clarity, we assume that $\textbf{Y}$ and $\bm{\Psi}$ are already centralized. Note that we pursue effective prediction of the scalar response rather than the estimation of $\bm{\beta}$ in (\ref{model_mat}).

The NIPALS-PLS algorithm is employed for (\ref{model_mat}). The algorithm tends to find two directions for $\bm{\Psi}$ and $\textbf{Y}$ respectively, such that the covariance of their projections on these two directions can be maximized. Then the projection of $\bm{\Psi}$ is used in the regression of $\textbf{Y}$. After deflation of both $\bm{\Psi}$ and $\textbf{Y}$, the second projection of $\bm{\Psi}$ can be obtained in the same way and is added as the second predictor for the regression of $\textbf{Y}$. The above procedure will repeat until $q$ projections are attained, where $q$ is the number of components and is pre-chosen in the computation. Let $w_l$, $c_l$ denote the $l$-th projection directions of $\bm{\Psi}$ and $\textbf{Y}$, and $t_l$ and $u_l$ denote the $l$-th projections of $\bm{\Psi}$ and $\textbf{Y}$. It is shown in \citet{rannar1994pls} that $w_l$, $c_l$, $t_l$ and $u_l$ are the $l$-th eigenvectors of $\bm{\Psi}^\top \textbf{Y} \textbf{Y}^\top \bm{\Psi}$, $\textbf{Y}^\top \bm{\Psi} \bm{\Psi}^\top \textbf{Y}$, $\bm{\Psi} \bm{\Psi}^\top \textbf{Y} \textbf{Y}^\top$ and $\textbf{Y} \textbf{Y}^\top \bm{\Psi} \bm{\Psi}^\top$ respectively. Specifically, the main procedure of the NIPALS algorithm can be summarized as follows \citep{rosipal2001kernel}:
\begin{itemize}

\item[(1)] Start with $l = 1$.
\item[(2)] Randomly initialize $u_l$.
\item[(3)] $t_l = \bm{\Psi} \bm{\Psi}^\top u_l$, $t_l \leftarrow t_l / \|t_l\|_2$, where $\|t_l\|_2 = (t_l^\top t_l)^{1/2}$.
\item[(4)] $u_l = \textbf{Y}\textbf{Y}^\top t_l$, $u_l \leftarrow u_l / \|u_l\|_2$, where $\|u_l\|_2 = (u_l^\top u_l)^{1/2}$.
\item[(5)] Repeat (3)-(4), until convergence.
\item[(6)] Deflation: $\bm{\Psi} \leftarrow \bm{\Psi} - t_lt_l^\top \bm{\Psi}$, $\textbf{Y} \leftarrow \textbf{Y} - t_lt_l^\top \textbf{Y}$. Then $l \leftarrow l + 1$.
\item[(7)] Repeat (2)-(6), until $l > q$.

\end{itemize}
Let $U = (u_1, \ldots, u_q)$ and $T = (t_1, \ldots, t_q)$. Then for $\wt{X}_1, \ldots, \wt{X}_n$ and a set of new observations $\wt{X}_{0, 1}, \ldots, \wt{X}_{0, m}$, we have
\begin{align}
\wh{\textbf{Y}} &= \bm{\Psi} \bm{\Psi}^\top U (T^\top \bm{\Psi} \bm{\Psi}^\top U)^{-1} T^\top \textbf{Y}, \nonumber \\
\wh{\textbf{Y}}_0 &= \bm{\Psi}_0 \bm{\Psi}^\top U (T^\top \bm{\Psi} \bm{\Psi}^\top U)^{-1} T^\top \textbf{Y}, \nonumber
\end{align}
where $\wh{\textbf{Y}}_0 = (\wh{Y}_{0, 1}, \ldots, \wh{Y}_{0, m})^\top$ is the prediction of the scalar response of the new observations, $\bm{\Psi}_0 = (\Psi_{0, 1}, \ldots, \Psi_{0, m})^\top$  and $\Psi_{0, i} = \bm{\Phi}(\wt{X}_{0, i})$.

\subsection{Kernel functional partial least squares}

It can be observed that $\bm{\Psi} \bm{\Psi}^\top$ is frequently appeared in the computation, but it can not be obtained directly since the nonlinear map $\bm{\Phi}$ is unknown. However, only inner product is involved in the computation of $\bm{\Psi} \bm{\Psi}^\top$, so kernel trick is applied. In specific, let $\bm{\Phi}(\wt{X}_i)^\top \bm{\Phi}(\wt{X}_h) = K(\wt{X}_i, \wt{X}_h)$, where $K( \cdot, \cdot)$ is the kernel function. Throughout the paper, Gaussian kernel is used, which is defined as
\begin{align}
K(\wt{X}_i, \wt{X}_h) = \exp \{ - \gamma \| \wt{X}_i - \wt{X}_h \|_{\mathcal{X}^p}^2 \}, \nonumber
\end{align}
where $\| \wt{X}_i - \wt{X}_h \|_{\mathcal{X}^p}^2 = \sum_{j = 1}^p \int_{\mathcal{T}} \{ X_{ij}(t) - X_{hj}(t) \}^2 dt$ and $\gamma$ is the tuning parameter. Moreover, \citet{song2021nonlinear} discussed the advantages of using Gaussian kernel in functional context. Denote $\textbf{K}$ as the Gram matrix whose $(i, h)$-th element equals to $K(\wt{X}_i, \wt{X}_h)$, and $\textbf{K}_0$ as the matrix whose $(i, h)$-th element equals to $K(\wt{X}_{0, i}, \wt{X}_h)$. Then the proposed KFPLS algorithm is as follows:
\begin{itemize}

\item[(1)] Input $\wt{X}_1, \ldots, \wt{X}_n$, $\textbf{Y}$, $\wt{X}_{0, 1}, \ldots, \wt{X}_{0, m}$. Start with $l = 1$.
\item[(2)] Randomly initialize $u_l$.
\item[(3)] $t_l = \textbf{K} u_l$, $t_l \leftarrow t_l / \|t_l\|_2$, where $\|t_l\|_2 = (t_l^\top t_l)^{1/2}$.
\item[(4)] $u_l = \textbf{Y}\textbf{Y}^\top t_l$, $u_l \leftarrow u_l / \|u_l\|_2$, where $\|u_l\|_2 = (u_l^\top u_l)^{1/2}$.
\item[(5)] Repeat (3)-(4), until convergence.
\item[(6)] Deflation: $\textbf{K} \leftarrow (I - t_lt_l^\top) \textbf{K} (I - t_lt_l^\top)$, $\textbf{Y} \leftarrow \textbf{Y} - t_lt_l^\top \textbf{Y}$. Then $l \leftarrow l + 1$.
\item[(7)] Repeat (2)-(6), until $l > q$.
\item[(8)] Output $\wh{\textbf{Y}} = \textbf{K} U (T^\top \textbf{K} U)^{-1} T^\top \textbf{Y}$ and $\wh{\textbf{Y}}_{0} = \textbf{K}_{0} U (T^\top \textbf{K} U)^{-1} T^\top \textbf{Y}$.

\end{itemize}

Note that the above algorithm is derived under the assumption that $\bm{\Psi}$ is centralized. Without this assumption, $\bm{\Psi}$ should be substituted by $\bm{\Psi} - \frac{1}{n} \textbf{1}_n \textbf{1}_n^\top \bm{\Psi}$, where $\textbf{1}_n = (1, \ldots, 1)^\top$ is a vector of length $n$. Further, we have
\begin{align}
\big (\bm{\Psi} - \frac{1}{n} \textbf{1}_n \textbf{1}_n^\top \bm{\Psi} \big ) \big (\bm{\Psi} - \frac{1}{n} \textbf{1}_n \textbf{1}_n^\top \bm{\Psi} \big )^\top &= \big ( \textbf{I} - \frac{1}{n} \textbf{1}_n \textbf{1}_n^\top \big ) \textbf{K} \big ( \textbf{I} - \frac{1}{n} \textbf{1}_n \textbf{1}_n^\top \big ) \triangleq \textbf{K}_c, \nonumber \\
\big (\bm{\Psi}_0 - \frac{1}{n} \textbf{1}_n \textbf{1}_n^\top \bm{\Psi} \big ) \big (\bm{\Psi} - \frac{1}{n} \textbf{1}_n \textbf{1}_n^\top \bm{\Psi} \big )^\top &= \big ( \textbf{K}_0 - \frac{1}{n} \textbf{1}_n \textbf{1}_n^\top \textbf{K} \big ) \big ( \textbf{I} - \frac{1}{n} \textbf{1}_n \textbf{1}_n^\top \big ) \triangleq \textbf{K}_{0c}. \nonumber
\end{align}
Therefore, $\textbf{K}$ and $\textbf{K}_0$ should be replaced by $\textbf{K}_c$ and $\textbf{K}_{0c}$ in the above KFPLS algorithm respectively.

We complete this section by discussing the selection of $q$ and $\gamma$, which are the number of components and the parameter of the kernel function respectively. The $V$-fold cross-validation method is applied. We randomly divide $(\wt{X}_1, Y_1), \ldots, (\wt{X}_n, Y_n)$ into $V$ parts. For the $v$-th part, the prediction of the scalar response is validated by the model that trained using the other $V - 1$ parts. Then the CV score can be obtained by
\begin{align}
\mbox{CV}(q, \gamma) = \frac{1}{V} \sum_{v = 1}^V \frac{1}{n_v} \sum_{i = 1}^{n_v} \big ( Y_{vi} - \wh{Y}_{vi} \big )^2,  \nonumber
\end{align}
where $n_v$ is the sample size of the $v$-th part, $Y_{vi}$ is the response of the $i$-th subject in the $v$-th part, and $\wh{Y}_{vi}$ is the prediction of $Y_{vi}$ when the tuning parameters are selected by $q$ and $\gamma$. The CV score is computed for a set of candidates of $q$ and $\gamma$ and the minimal CV score is preferred.

\section{ Simulation }\label{SecSim}

In this section, simulation studies are conducted to assess the performance of the proposed KFPLS method. For comparison, we also consider the classical FLR method introduced in \citep{ramsay2005functional} and the approach developed for FGAM in \citep{mclean2014functional}. For simplicity, these two methods are denoted as FLR and FGAM respectively. Further, various functional nonlinear models are taken into account when generating the simulated data, and the detailed settings are summarized as follows:
\begin{itemize}

\item Scenario 1: We generate the data via the following functional nonlinear model,
\begin{align}
Y_i = g \Big (\int_{\mathcal{T}} X_{i1}(t) \beta_1(t) dt + \int_{\mathcal{T}} X_{i2}(t) \beta_2(t) dt \Big ) + \epsilon_i, \nonumber
\end{align}
where $g$ is a nonparametric function, $\beta_1(t)$ and $\beta_2(t)$ are the coefficient functions. The simulated data for the covariate functions are obtained through $X_{i1}(t) = \sum_{l} a_{il} B_l(t)$ and $X_{i2}(t) = \sum_{l} b_{il} B_l(t)$, where $a_{il}$ and $b_{il}$ are generated by the standard normal distribution, and $B_l(t)$ is the $l$-th B-spline basis function with order $4$ and $21$ knots. Moreover, the measurement errors are generated from $N(0, 0.05^2)$. We set $\beta_1(t) = 2 \cdot \sin (2 \pi t)$, $\beta_2(t) = 2 \cdot \cos (2 \pi t)$ and $\mathcal{T} = [0, 1]$. For the nonparametric function, we also consider different setups: (1) Case 1: $g(x) = x$; (2) Case 2: $g(x) = x^2$; (3) Case 3: $g(x) = \cos(\pi x / 2)$; (4) Case 4: $g(x) = 10 \cdot \sin(x) / x$.

\item Scenario 2: We generate the data via the following functional nonlinear model,
\begin{align}
Y_i = g_1 \Big (\int_{\mathcal{T}} X_{i1}(t) \beta_1(t) dt \Big ) + g_2 \Big ( \int_{\mathcal{T}} X_{i2}(t) \beta_2(t) dt \Big ) + \epsilon_i, \nonumber
\end{align}
where $g_1$ and $g_2$ are distinct nonparametric function, $\beta_1(t)$ and $\beta_2(t)$ are the coefficient functions. The simulated data for the covariate functions $X_{i1}$ and $X_{i2}$, and the measurement error $\epsilon_i$ are obtained in the same way as Scenario 1. Moreover, the settings of $\beta_1(t)$, $\beta_2(t)$ and $\mathcal{T}$ are the same as that of Scenario 1. For the nonparametric functions, we also consider different setups: (1) Case 1: $g_1(x) = \cos(\pi x / 2)$ and $g_2(x) = \sin(\pi x / 2)$; (2) Case 2: $g_1(x) = \cos(\pi x / 2)$ and $g_2(x) = x^2$.

\item Scenario 3: We generate the data via the following functional nonlinear model,
\begin{align}
Y_i = \int_{\mathcal{T}} F_1\{X_{i1}(t), t\} dt + \int_{\mathcal{T}} F_2\{X_{i2}(t), t\} dt + \epsilon_i, \nonumber
\end{align}
where $F_1$ and $F_2$ are two smooth functions. The simulated data for the covariate functions $X_{i1}$ and $X_{i2}$, and the measurement error $\epsilon_i$ are obtained in the same way as Scenario 1. Moreover, we set $\mathcal{T} = [0, 1]$ as Scenario 1. For the smooth function, we set $F_1(x, t) = t \cdot \sin(x)$ and $F_2(x, t) = t \cdot \cos(x)$.

\end{itemize}
By considering the above scenarios, we can explore the performance of our method when facing data from different nonlinear models. Note that the model for Case 1 of Scenario 1 is exactly a functional linear regression model with two functional predictors, as the nonparametric function is set as an identity function. Furthermore, the model considered in Scenario 3 is the functional generalized additive model introduced in \citet{mclean2014functional}.

The simulated data are split into training set and test set, which are used for model fitting and performance assessment respectively. The size of training set is set as $n = 200$ and $n = 500$, and the size of test set is $500$. We evaluate the quality of the scalar response prediction by the root average squared error (RASE) and the average relative prediction error (ARPE), which are defined as
\begin{align}
\mbox{RASE} &= \sqrt{ \wt{n}^{-1} \sum_{i = 1}^{\wt{n}} \big (\wh{Y}_i - Y_i \big )^2 }, \nonumber \\
\mbox{ARPE} &= \wt{n}^{-1} \sum_{i = 1}^{\wt{n}} \frac{|\wh{Y}_i - Y_i|}{|Y_{max}|}, \nonumber
\end{align}
where $Y_{max} = \max_{i} |Y_i|$ and $\wt{n}$ is the sample size of the training set or test set. Here we use $Y_{max}$ in the computation of $\mbox{ARPE}$, as the small value of $Y_i$ will make the relative prediction error extremely large and the ARPE can be quite unstable. We conduct 100 Monte Carlo runs for each case, and RASE and ARPE are computed for each run.

Table \ref{SimTabScen1} lists the simulation results of KFPLS, FLM and FGAM methods in Scenario 1, and average RASE and ARPE of both training set and test set are presented. For Case 1, all three methods display similar prediction ability for the scalar response, as linear model is used for data generation in this case. However, FLM and FGAM methods show obvious deterioration for both RASE and ARPE in Case 2, since the simulated data cannot satisfy the model assumption of these two methods. Furthermore, the increasing sample size of training set from $200$ to $500$ do not bring apparent help in the prediction of scalar response in the test set. On the contrary, the proposed KFPLS method performs the best for both training set and test set in Case 2. As the size of training set becomes lager, the prediction error of KFPLS method in the test set decreases accordingly. Moreover, results of Case 3 exhibit similar trend as that of Case 2, and our KFPLS method still gives the most accurate prediction for both training set and test set. In Case 4, KFPLS method also yields much smaller RASE and ARPE than FLM and FGAM methods. The above results indicates the superiority of the proposed KFPLS method in Scenario 1.

The average RASE and ARPE of KFPLS, FLM and FGAM methods in Scenario 2 are reported in Table \ref{SimTabScen2}. As the model used for data generation in Scenario 2 is not consistent with the model assumption of FLM and FGAM methods, it is evident that the simulation results of these two methods are not valid enough. However, our KFPLS method is still able to give more precise prediction in both cases of Scenario 2, regardless of whether in training set or test set. That also implies the advantages of our KFPLS method in handling functional nonlinear models. Further, Table \ref{SimTabScen3} provides the simulation results of Scenario 3. It can be observed that KFPLS and FGAM methods show minor difference with each other and perform better than FLM method in the training set. For the test set, the prediction errors of KFPLS method are slightly higher when compared with the errors of FGAM method. The reason for the good performance of FGAM method is that the simulated data in Scenario 3 can satisfy the model assumption of FGAM method. Nevertheless, the prediction results of the proposed KFPLS method is also reasonable in this scenario.

To sum up, the proposed KFPLS method shows encouraging prediction ability in all these three considered scenarios, which demonstrates the flexibility of our method to various functional nonlinear models. For the simulation results in all three scenarios, it is interesting to find that the prediction error of KFPLS in training set becomes higher with the rise of $n$, while the prediction error in test set can be smaller as $n$ increases. We conjecture the reason is that the method may be likely to face overfitting issue when only a small number of samples are available.

\begin{table}[htbp]
\caption{Average RASE and ARPE across 100 Monte Carlo runs in Scenario 1, with standard deviation in parentheses.}
\label{SimTabScen1}
\begin{center}
\setlength{\tabcolsep}{1mm}{
\begin{tabular}{ccccccc}
\hline
 & & & \multicolumn{2}{c}{Training set} & \multicolumn{2}{c}{Test set} \\
 & & & RASE & ARPE & RASE & ARPE \\
\hline
\multirow{6}{*}{Case 1}&\multirow{3}{*}{$n = 200$}& KFPLS &0.0472 (0.0027)&0.0304 (0.0041)&0.0552 (0.0023)&0.0320 (0.0037)\\
& & FLM &0.0532 (0.0083)&0.0345 (0.0074)&0.0613 (0.0101)&0.0359 (0.0078)\\
& & FGAM &0.0470 (0.0038)&0.0303 (0.0047)&0.0543 (0.0039)&0.0315 (0.0039)\\
\cline{2-7}
 &\multirow{3}{*}{$n = 500$}& KFPLS &0.0496 (0.0015)&0.0286 (0.0029)&0.0524 (0.0016)&0.0307 (0.0033)\\
& & FLM &0.0527 (0.0043)&0.0305 (0.0043)&0.0556 (0.0045)&0.0326 (0.0045)\\
& & FGAM &0.0494 (0.0017)&0.0285 (0.0030)&0.0522 (0.0019)&0.0306 (0.0033)\\
\hline
\multirow{6}{*}{Case 2}&\multirow{3}{*}{$n = 200$}& KFPLS &0.0503 (0.0178)&0.0249 (0.0085)&0.1873 (0.0212)&0.0712 (0.0151)\\
& & FLM &0.2390 (0.0291)&0.1102 (0.0223)&0.2799 (0.0236)&0.1064 (0.0232)\\
& & FGAM &0.2140 (0.0294)&0.1004 (0.0240)&0.2752 (0.0282)&0.1034 (0.0229)\\
\cline{2-7}
 &\multirow{3}{*}{$n = 500$}& KFPLS &0.0636 (0.0140)&0.0268 (0.0063)&0.1383 (0.0126)&0.0523 (0.0104)\\
& & FLM &0.2491 (0.0177)&0.0939 (0.0190)&0.2696 (0.0219)&0.0991 (0.0190)\\
& & FGAM &0.2265 (0.0167)&0.0870 (0.0187)&0.2644 (0.0255)&0.0967 (0.0183)\\
\hline
\multirow{6}{*}{Case 3}&\multirow{3}{*}{$n = 200$}& KFPLS &0.0503 (0.0179)&0.0359 (0.0129)&0.1872 (0.0144)&0.1243 (0.0083)\\
& & FLM &0.2398 (0.0207)&0.1620 (0.0152)&0.2830 (0.0171)&0.1850 (0.0107)\\
& & FGAM &0.2151 (0.0239)&0.1466 (0.0165)&0.2750 (0.0233)&0.1779 (0.0139)\\
\cline{2-7}
 &\multirow{3}{*}{$n = 500$}& KFPLS &0.0604 (0.0146)&0.0426 (0.0104)&0.1355 (0.0086)&0.0912 (0.0051)\\
& & FLM &0.2518 (0.0144)&0.1652 (0.0096)&0.2686 (0.0140)&0.1756 (0.0079)\\
& & FGAM &0.2310 (0.0130)&0.1526 (0.0091)&0.2595 (0.0144)&0.1691 (0.0086)\\
\hline
\multirow{6}{*}{Case 4}&\multirow{3}{*}{$n = 200$}& KFPLS &0.0742 (0.0267)&0.0058 (0.0021)&0.2841 (0.0242)&0.0197 (0.0012)\\
& & FLM &0.3759 (0.0459)&0.0264 (0.0028)&0.4348 (0.0343)&0.0302 (0.0023)\\
& & FGAM &0.3299 (0.0425)&0.0235 (0.0027)&0.4339 (0.0677)&0.0296 (0.0036)\\
\cline{2-7}
 &\multirow{3}{*}{$n = 500$}& KFPLS &0.0912 (0.0221)&0.0071 (0.0017)&0.2046 (0.0197)&0.0143 (0.0011)\\
& & FLM &0.3887 (0.0300)&0.0271 (0.0018)&0.4124 (0.0293)&0.0287 (0.0014)\\
& & FGAM &0.3536 (0.0258)&0.0250 (0.0016)&0.4013 (0.0293)&0.0278 (0.0015)\\
\hline
\end{tabular}}
\end{center}
\end{table}

\begin{table}[htbp]
\caption{Average RASE and ARPE across 100 Monte Carlo runs in Scenario 2, with standard deviation in parentheses.}
\label{SimTabScen2}
\begin{center}
\setlength{\tabcolsep}{1mm}{
\begin{tabular}{ccccccc}
\hline
 & & & \multicolumn{2}{c}{Training set} & \multicolumn{2}{c}{Test set} \\
 & & & RASE & ARPE & RASE & ARPE \\
\hline
\multirow{6}{*}{Case 1}&\multirow{3}{*}{$n = 200$}& KFPLS &0.0562 (0.0064)&0.0234 (0.0028)&0.1336 (0.0101)&0.0518 (0.0037)\\
& & FLM &0.1536 (0.0129)&0.0595 (0.0068)&0.1776 (0.0120)&0.0671 (0.0054)\\
& & FGAM &0.1254 (0.0127)&0.0497 (0.0054)&0.1643 (0.0170)&0.0619 (0.0061)\\
\cline{2-7}
 &\multirow{3}{*}{$n = 500$}& KFPLS &0.0672 (0.0031)&0.0273 (0.0015)&0.1086 (0.0064)&0.0424 (0.0026)\\
& & FLM &0.1573 (0.0091)&0.0582 (0.0043)&0.1687 (0.0104)&0.0620 (0.0041)\\
& & FGAM &0.1357 (0.0072)&0.0518 (0.0029)&0.1532 (0.0092)&0.0574 (0.0029)\\
\hline
\multirow{6}{*}{Case 2}&\multirow{3}{*}{$n = 200$}& KFPLS &0.0383 (0.0141)&0.0178 (0.0067)&0.1467 (0.0093)&0.0592 (0.0064)\\
& & FLM &0.1819 (0.0141)&0.0774 (0.0078)&0.2125 (0.0126)&0.0840 (0.0096)\\
& & FGAM &0.1507 (0.0137)&0.0666 (0.0080)&0.1990 (0.0167)&0.0803 (0.0104)\\
\cline{2-7}
 &\multirow{3}{*}{$n = 500$}& KFPLS &0.0516 (0.0114)&0.0220 (0.0048)&0.1087 (0.0079)&0.0438 (0.0050)\\
& & FLM &0.1914 (0.0109)&0.0722 (0.0063)&0.2048 (0.0103)&0.0778 (0.0085)\\
& & FGAM &0.1635 (0.0088)&0.0654 (0.0059)&0.1876 (0.0106)&0.0746 (0.0082)\\
\hline
\end{tabular}}
\end{center}
\end{table}

\begin{table}[htbp]
\caption{Average RASE and ARPE across 100 Monte Carlo runs in Scenario 3, with standard deviation in parentheses.}
\label{SimTabScen3}
\begin{center}
\setlength{\tabcolsep}{1mm}{
\begin{tabular}{cccccc}
\hline
 & & \multicolumn{2}{c}{Training set} & \multicolumn{2}{c}{Test set} \\
 & & RASE & RPE & RASE & RPE \\
\hline
\multirow{3}{*}{$n = 200$}& KFPLS &0.0413 (0.0108)&0.0462 (0.0121)&0.0665 (0.0028)&0.0727 (0.0050)\\
& FLM &0.0614 (0.0037)&0.0694 (0.0061)&0.0714 (0.0030)&0.0781 (0.0052)\\
& FGAM &0.0476 (0.0037)&0.0537 (0.0051)&0.0546 (0.0024)&0.0598 (0.0041)\\
\hline
\multirow{3}{*}{$n = 500$}& KFPLS &0.0488 (0.0069)&0.0531 (0.0083)&0.0623 (0.0021)&0.0678 (0.0039)\\
& FLM &0.0646 (0.0021)&0.0704 (0.0046)&0.0677 (0.0022)&0.0736 (0.0041)\\
& FGAM &0.0503 (0.0020)&0.0549 (0.0036)&0.0523 (0.0020)&0.0570 (0.0035)\\
\hline
\end{tabular}}
\end{center}
\end{table}

\section{ Real Data Analysis }\label{SecReal}

\begin{figure}[H]
  \centering
  \includegraphics[width=0.6\textwidth]{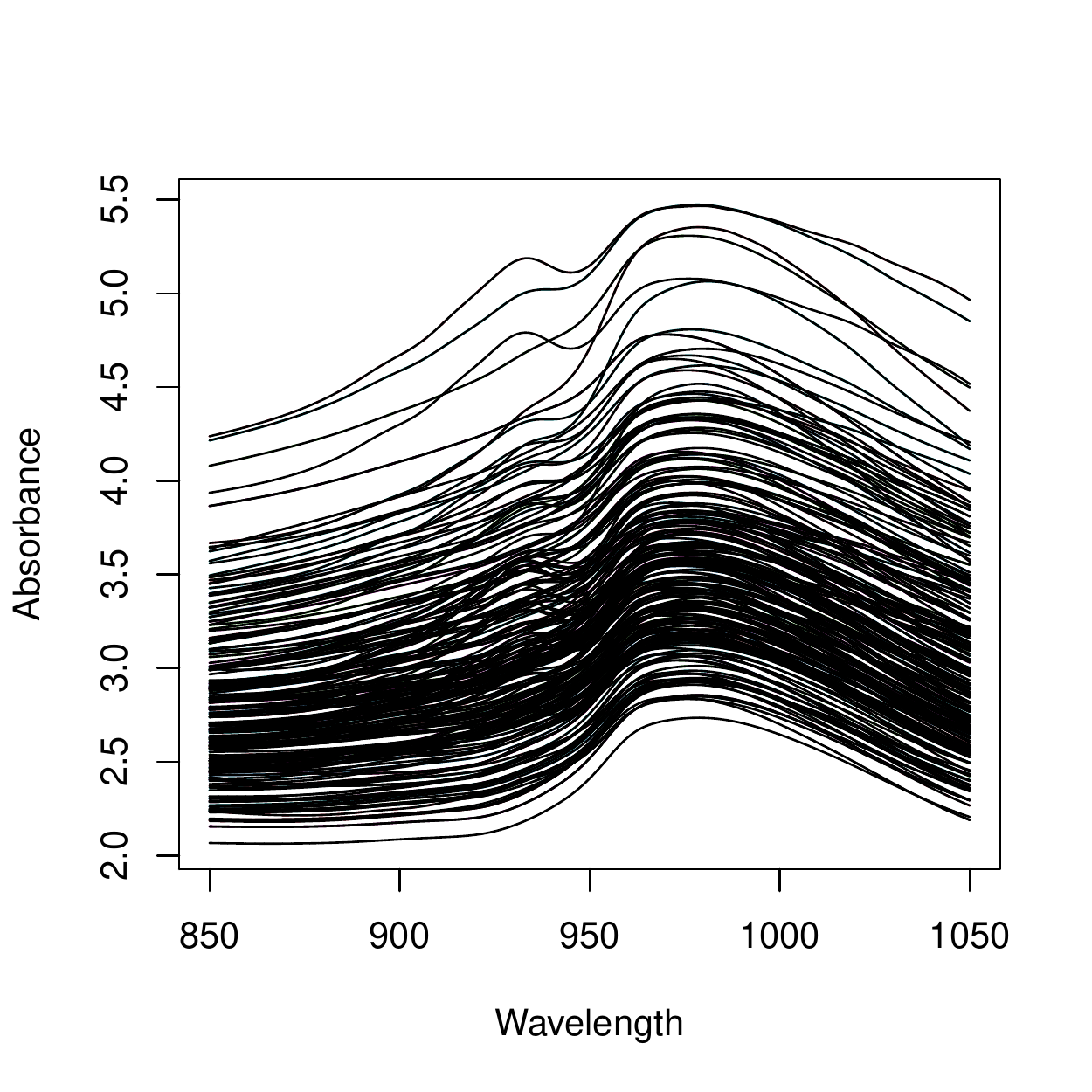}\\
  \caption{The spectrometric curves of the $215$ meat samples.}
  \label{Figreal}
\end{figure}

In this section, we further demonstrate the superiority of the proposed KFPLS method through real application. The Tecator dataset is taken into account, and the data can be obtained from \url{https://www.math.univ-toulouse.fr/~ferraty/SOFTWARES/NPFDA/npfda-datasets.html}. There are $215$ pieces of finely chopped meat considered in this study. For each meat sample, a spectrometric curve that corresponds to the absorbance measured at $100$ wavelengths from $850$ nm to $1050$ nm is observed, and the fat content is also recorded. Moreover, the spectrometric curves of these $215$ meat samples are shown in Figure \ref{Figreal}. We aim to achieve precise prediction of the fat content based on the spectrometric curve.

For implementation of our KFPLS method, the fat content is treated as the scalar response, and the spectrometric curve is treated as the functional predictor. Furthermore, we transform the wavelength interval to $[0, 1]$. The dataset is randomly divided into training set and test set, with the size of $165$ and $50$ respectively. As a comparison, the FLM and FGAM methods are also applied to the dataset. We randomly split the dataset for 100 times and evaluate these three methods via ARSE and ARPE that defined in Section \ref{SecSim}.

Table \ref{RealTab} reports the average ARSE and ARPE for KFPLS, FLM and FGAM methods over 100 runs. It is shown that these three methods perform similarly for the training set. Nevertheless, it is evident that the prediction results of FLM method become invalid for the test set, which indicates that the linear assumption may not be satisfied for the Tecator dataset. Moreover, the RASE and ARPE of FGAM approach are also higher than that of KFPLS method, with greater standard deviation. The proposed KFPLS method works well for both training set and test set. The promising performance of KFPLS method implies that the proposed method is more flexible and able to give accurate prediction for more types of nonlinear models.

\begin{table}[htbp]
\caption{Average RASE and ARPE of KFPLS, FLM and FGAM methods across 100 runs for the Tecator dataset, with standard deviation in parentheses.}
\label{RealTab}
\begin{center}
\setlength{\tabcolsep}{1mm}{
\begin{tabular}{ccccc}
\hline
 & \multicolumn{2}{c}{Training set} & \multicolumn{2}{c}{Test set} \\
 & RASE & RPE & RASE & RPE \\
\hline
KFPLS &2.3265 (0.1169)&0.0356 (0.0020)&2.6294 (0.3651)&0.0411 (0.0052)\\
FLM &2.4956 (0.0952)&0.0412 (0.0018)&168.9224 (175.3095)&3.5244 (3.6547)\\
FGAM &2.1715 (0.1056)&0.0325 (0.0018)&4.1604 (7.5939)&0.0482 (0.0315)\\
\hline
\end{tabular}}
\end{center}
\end{table}

\section{ Conclusion and Discussion }\label{SecDis}

In this paper, we focus on functional nonlinear models with scalar response and functional predictors. As the existing approaches still require particular assumptions on the structure of the nonlinear relationship, more flexibility of the nonlinear models can be quite appealing. Inspired by the kernel method in machine learning, we utilize the corresponding idea in the framework of functional regression and introduce the KFPLS method. The proposed method can handle models with multiple functional predictors and do not need strict limits on the nonlinear structure. The results of the simulation study illustrate that the KFPLS method can adapt to various types of nonlinear models. Furthermore, the analysis of the Tecator dataset indicates the encouraging predicting ability of the KFPLS method in real world application.

There are also some possible extensions can be considered for the current work. First, we only consider functional predictors for the functional nonlinear models. Many existing works showed interests in the functional regression models with both functional predictors and scalar predictors, such as \citet{wong2019partially} and \citet{tang2023estimation}. Hence, it can be useful to extend our method to functional nonlinear model with both types of predictors. Second, though multiple predictors are available for the considered model, variable selection has to be conducted when the number of predictors is large. Therefore, developing new variable selection approach for functional regression and combining it with our algorithm can be helpful.

\section*{Acknowledgements}

This work was supported by Public Health $\&$ Disease Control and Prevention, Major Innovation $\&$ Planning Interdisciplinary Platform for the ``Double-First Class" Initiative, Renmin University of China. This work was supported by the Outstanding Innovative Talents Cultivation Funded Programs 2021 of Renmin University of China.

\bibliographystyle{unsrtnat}
\bibliography{ref}

\end{document}